# Variable Electrical Responses in Epitaxial Graphene Nanoribbons

C.-C. Yeh[1,2†], S. M. Mhatre[1,3†], N. T. M. Tran[1,4], H. M. Hill[1], H. Jin[1,4], P.-C. Liao[3], D. K. Patel[2,5], R. E. Elmquist[1], C.-T. Liang[2], and A. F. Rigosi[1,a]

[1]*Physical Measurement Laboratory, National Institute of Standards and Technology (NIST), Gaithersburg, Maryland, 20899-8171, USA*

[2]*Department of Physics, National Taiwan University, Taipei, 10617, Taiwan*

[3]*Graduate Institute of Applied Physics, National Taiwan University, Taipei 10617, Taiwan*

[4]*Joint Quantum Institute, University of Maryland, College Park, MD 20742, USA*

[5]*Electricity Division, Physikalisch-Technische Bundesanstalt (PTB), 38116 Brunswick, Germany*

We have demonstrated the fabrication of both armchair and zigzag epitaxial graphene nanoribbon (GNR) devices on 4H-SiC using a polymer-assisted sublimation growth method. The phenomenon of terrace step formation has traditionally introduced the risk of GNR deformation along sidewalls, but a polymer-assisted sublimation method helps mitigate this risk. Each type of 50 nm wide GNR is examined electrically and optically (armchair and zigzag), with the latter method being a check on the quality of the GNR devices and the former using alternating current to investigate resistance attenuation from frequencies above 100 Hz. Rates of attenuation are determined for each type of GNR device, revealing subtle suggested differences between armchair and zigzag GNRs.

___________________________

[a)] Author to whom correspondence should be addressed.  Email: afr1@nist.gov

[b)] †These authors contributed equally to this work

Graphene nanoribbons (GNRs) have been shown to harbor desirable electrical properties due to their unique band structure [1-3]. Even though calculations of relevant band structures have been studied in depth, more experimental work is needed to verify the electrical behaviors that distinguish armchair (AC) GNRs [4-8] and zigzag (ZZ) GNRs [9-12]. In the former case, the edge segments are rotated by 120° relative to the previous edge segment, whereas the latter case have edge segments that alternate in their orientation with respect to the direction along the edge. An advantageous quality of GNRs is that one can observe quantum confinement effects [13-20], and this typically manifests as a change in band gap energies that are inversely proportional to the width of the ribbon. Because varying band gaps are applicable to many engineering pursuits, it is still crucial that one understands that ongoing complications that still exist with regard to controlling ribbon width naturally as opposed to doing so with fabrication methods. Thus, continued efforts are warranted in order to fully comprehend epitaxially grown GNRs.

As mentioned above, methods that involve lithographic patterning typically cause disordered edges to form as a result of the etching process. This disorder contributes to the degradation of the GNR's electrical properties, as shown in other works [21-29]. Avoiding this degradation in structural quality is possible with silicon carbide (SiC) substrate, which may provide a suitable template for GNR growth specific to various crystal orientations. Epitaxial GNRs have exhibited excellent ballistic transport characteristics and electronic mean free paths of up to 15 μm [30-41]. Despite having avoided the issue of the lithographic disorder, there are problems exclusive to these growths that need to be taken into account, like GNR distortion resulting from terrace formations [29]. Though it is possible to avoid these terrace effects via device fabrication from self-assembled GNRs on SiC [42], device sizes then become constrained to the size of the terrace. Using polymer-assisted sublimation growth (PASG) techniques has enabled high-quality graphene growth on centimeter scales by suppressing terrace formation [43-47], rendering that graphene crucial for resistance metrology [48-52]. Therefore, such a technique could be used to better facilitate GNR growth.

In this work, a focus is placed on examining the electrical properties of both AC and ZZ GNR devices (via magnetotransport), which were measured at low temperatures. In addition to basic electrical transport properties, an investigation of the frequency-dependent resistance behaviors was also conducted to clarify the differences that may have been consistently present between AC and ZZ GNR devices. Additional techniques were also used to verify the structural quality include atomic force microscopy (AFM), conductive AFM, and Raman spectroscopy, with data from the lattermost technique suggesting a greater strain present in ZZ GNRs than in the AC GNRs [53]. the growths of approximately 50 nm-wide GNR devices are demonstrated with improved structural properties using PASG techniques. Because PASG helps prevent deformation of the necessary slanted SiC sidewalls during growth, GNRs are more likely to remain confined to the patterned substrate.

There are three main segments to the experiment: (1) fabrication, (2) verification, and (3) measurement. For the first segment, GNR samples are grown on square SiC chips, which are first diced from on-axis 4H-SiC(0001) semi-insulating wafers (CREE) [see notes]. Chips are then cleaned with piranha solution to remove organic contaminants, and any remaining contamination is easily removed by the solution of hydrofluoric acid. Two separate but consecutive etch processes were performed to shape the substrate in a way that would promote sidewall GNR growth (see Supplementary Material). Just before the growth process is initiated, chips were processed with a photoresist compound in solution (AZ5214E) for PASG [43]. The growth was performed with a graphite-lined resistive-element furnace (Materials Research Furnaces Inc.) [see notes], in an inert argon environment, and with heating and cooling rates of about 1.5 °C/s. The formation of GNRs on SiC occurred at 1400 °C for the 25 min duration of the furnace activity, with GNRs forming along both the $(1\bar{1}00)$ and $(11\bar{2}0)$ crystallographic directions on SiC. These orientations represent the ZZ and AC orientations, respectively. After the growth, the final chips are removed and then loaded into an electron beam deposition chamber. Layers of Pd and Au were deposited on the GNRs to protect the surface against organic contamination as excess GNRs were etched away. This protective layer was also useful during fabrication processes, where contact pads made of Pt, Au, and Ti were deposited. Further details are provided in the Supplementary Material.

The fabrication process is summarized by illustration in Fig. 1 (a). The GNRs of interest are centered in Fig. 1 (b) and (c), and due to their size, fabrication processes for depositing electrical contacts were carefully tested beforehand. These images were obtained via confocal laser scanning microscope (CLSM) with an Olympus LEXT OLS4100 system that was operated in an argon environment and fitted with objectives ranging from 5× to 100× and software features allowing an additional 8× optical zoom. The system utilizes a 405 nm wavelength semiconductor laser, which is scanned in the *x-y* directions by an electromagnetic micro-electro-mechanical systems (MEMS) scanner and a high-precision Galvano mirror. The use of CLSM for graphene on SiC and other low-dimensional materials has been well-described in other work [54]. The primary reason for using this technique, despite its ability to not directly see the GNRs (since its lateral resolution is about 150 nm), is for determining the extent to which buffer layer material (that is, a carbon layer that is covalently bonded to the SiC substrate) has been partially or fully grown in the surrounding areas, as this could be indicative of unsuitable GNRs.

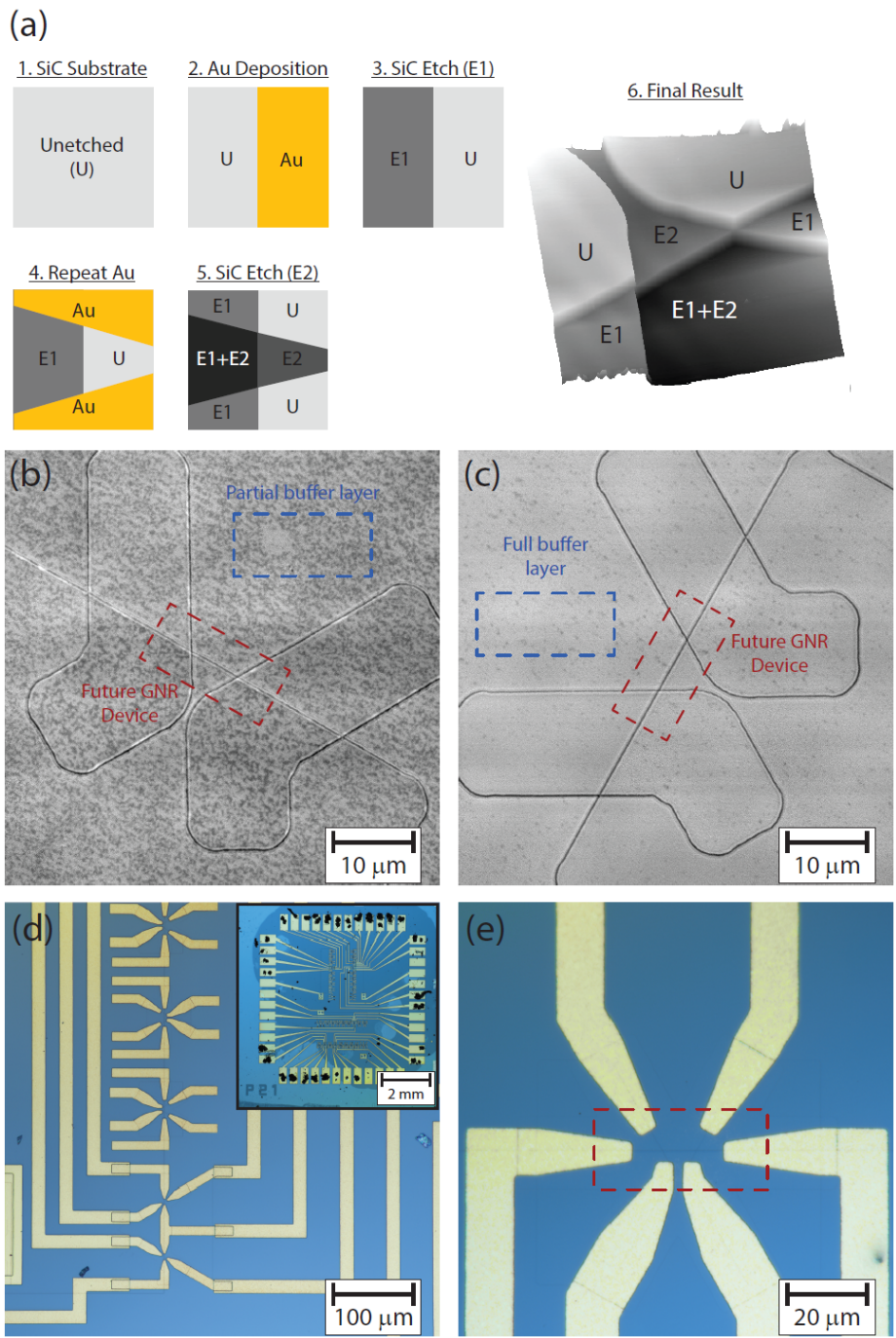

FIG. 1. (a) The GNR fabrication process is illustrated in a few steps. (b) ZZ and (c) AC GNR CLSM images are shown to exemplify that, in conjunction with later data, minor differences in surrounding growth do not indicate a loss of GNR quality. The ZZ GNR has less surrounding buffer layer, resulting in darker patches along the SiC surface. The GNRs of interest are in the center of the image and are the connecting element to the butterfly-like structure resulting from the two etching processes. (d) ZZ and (e) AC GNR optical images show the final devices, with the former including an inset of the whole chip. Both types of GNR can coexist when rotated by 90°. Optical images were acquired with a Nikon L200N microscope operating in reflection mode using a white light source [see notes].

Fortunately, the primary flat of the full SiC wafer provided an adequate crystallographic orientation to within 1°, allowing both types of GNR to be developed on the same chip with a simple 90° rotation. The final resulting devices appear in Fig. 1 (d) and (e) for ZZ and AC GNRs, respectively. Though faint, the lines demarcating the sidewalls upon which the GNRs were grown are visible optically by eye. The inset of Fig. 1 (d) shows the full device, with the top half hosting ZZ GNR devices and the bottom half hosting AC GNR devices. Images showing the end results of the substrate etching processes are shown in Fig. 1. The CLSM images (Fig. 1 (b) and (c)) show the ZZ and AC GNRs, respectively. There are minor differences in appearance, which are attributable to slightly different growth rates at different locations along the chip. In Fig. 1 (b), the darker, patchy appearance arises from the lack of graphene grown on the substrate, whereas Fig. 1 (c) shows a more, nearly complete, homogeneous film. In both cases, however, the GNR has grown properly, as concluded by the data in Fig. 2 in the second segment.

For the second segment, verification was required to ensure the growth quality and the device functionality. To collect AFM and C-AFM images, an Asylum Cypher system was used in contact mode with a Cr/Pt probe of radius 25 nm [see notes]. The C-AFM scans were performed with a bias voltage of 10 mV applied to the sample. The gain parameter was set to 10 and the setpoint varied between 0.2 V and 0.5 V. The C-AFM images in particular were more easily acquired due to the adequate electrical conductivity and adhesion exhibited by the GNRs, ultimately allowing one to accurately map the nanostructures. A more explicit profile of an example GNR device is shown in Fig. 2 (c) to show direct evidence for post-growth GNR width determination (approximately 50 nm).

To gauge the quality of the material, Raman spectra were collected using a Renishaw InVia micro-Raman spectrometer and a 633 nm wavelength excitation laser source [see notes]. Measurements were performed while utilizing a backscattering configuration with the sample upside-down, a pose that enhances the GNRs' optical responses while also reducing the contributions of the SiC optical response [55]. Other measurement parameters include a 2 μm spot size, 50 × objective, 1.7 mW power, 240 s acquisition time, and 1800 mm$^{-1}$ grating. Each GNR sample had its own Raman line map measurement, meaning that spectra were collected along the dimension of the GNR and numbered at least five in step sizes of about 1 μm.

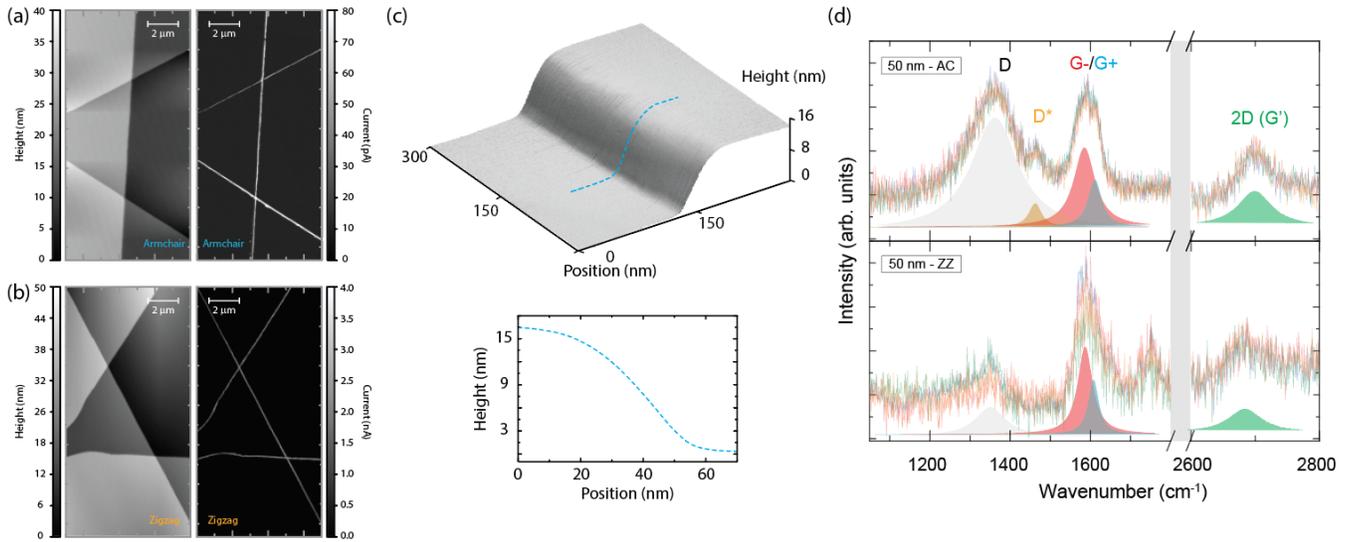

FIG. 2. (a) Standard AFM (left) and C-AFM (right) images are shown for an example 50 nm-wide GNR on 4H-SiC. The AC GNR exhibits a lower conductivity than its ZZ counterpart in (b). (b) The same type of data are shown for an example ZZ GNR. These techniques, especially C-AFM, are advantageous in that electrical quality can be determined in a nondestructive manner. One key feature to note is the continuity of electrical current that passes through the GNR along the entire device region, indicating sufficient growth. (c) A more explicit profile of an example GNR device is shown as direct evidence for post-growth GNR width determination (approximately 50 nm). (d) Raman spectra are shown. The top and bottom panels display 50 nm AC and 50 nm ZZ GNRs, respectively. Each panel contains five example spectra taken along the GNR. Consistently appearing modes, fit to Lorentzian profiles, include the D, D*, G-, G+, and 2D (G') bands.

To reassure the overall material quality of the GNRs, their final devices were tested for electrical functionality with C-AFM and imaged more generally with AFM. These images were acquired on both types of GNR devices of 50 nm width. In Fig. 2 (a) and (b), both standard AFM and C-AFM data are plotted for the AC and ZZ orientations, respectively. Note that ZZ GNRs exhibited higher conductivity than their AC counterparts by nearly two orders of magnitude. For additional support in determining GNR responsivity, Raman spectroscopy was performed with some spectra shown in Fig. 2 (c). The top and bottom panels display 50 nm AC and 50 nm ZZ GNRs, respectively. At least 30 spectra were acquired for each type of GNR, but only five example spectra per GNR type are shown in Fig. 2 (c) for visual clarity.

It is important to make a few remarks about the GNR Raman spectra. First, the G mode appears to split depending on the GNR type. This splitting has been recognized as a manifestation of strain present in the GNRs and varies based on GNR type and doping, with the latter being less relevant since epitaxially grown GNRs have been known to be charge neutral [53, 56-57]. Second, the 2D (G') mode behaves consistently with expectations from a GNR experiencing strain [53, 56-57], noting that ZZ-GNRs appear more redshifted. Lorentzian profiles were used to fit all spectra.

With characterization data verifying the devices' functionality, a more nuanced electrical transport assessment was then conducted using alternating current. Electrical characterization of the GNRs was performed using SR830 lock-in amplifiers and standard techniques from Stanford Research Systems [see notes]. Amplifiers were operated at several frequencies (alternating current) to obtain information on frequency-dependent resistances. A Keithley 2612A source measure unit was used to provide a direct current gate voltage. A CS580 voltage controlled current source was used to provide the source-drain current. A lock-in amplifier supplied the CS580 with a constant voltage of 1 V that it would then convert to a constant 1 µA of current for most measurements. For the AC GNRs, a voltage of 0.5 V, corresponding to a 0.5 µA was used for the longitudinal measurements. The magnetoresistance was measured in a Janis cryostat at a temperature of 4.4 K with a magnet capable of going up to ± 9 T at a rate of 0.01 T/s [see notes].

While holding the GNRs at various temperatures, both the longitudinal ($R_{xx}$) and orthogonal ($R_{xy}$) resistances were measured as a function of frequency. Electrical currents of 0.5 µA, 0.7 µA, and 1.0 µA were used for all the measurements, and symmetrical data also taken since each $R_{xx}$ and $R_{xy}$ had two measurable pairs. The resulting data are shown in Fig. 3 for the AC GNR devices and Fig. 4 for the ZZ GNR devices. Also included in each figure is an illustration of the GNR device and its markers used in these measurements.

In the panel array of Fig. 3 (b) (and Fig. 4 (b)), semi-logarithmic plots show the behavior of the two resistances ($R_{xx}$, left column, and $R_{xy}$, right column) as a function of frequency, with each panel showing data taken at either 293 K, 80 K, or 4.3 K. By inspecting each plot, a few consistencies can be described. First, three curves are shown per side of device, either in blue and red for top and bottom, respectively, (in the $R_{xx}$ case), or blue and red for left and right pair, respectively, (in the $R_{xy}$ case). Second, in some cases of very low frequency (near 10 Hz and below), saturation appears in some data as a result of the lock-in amplifier's limitations.

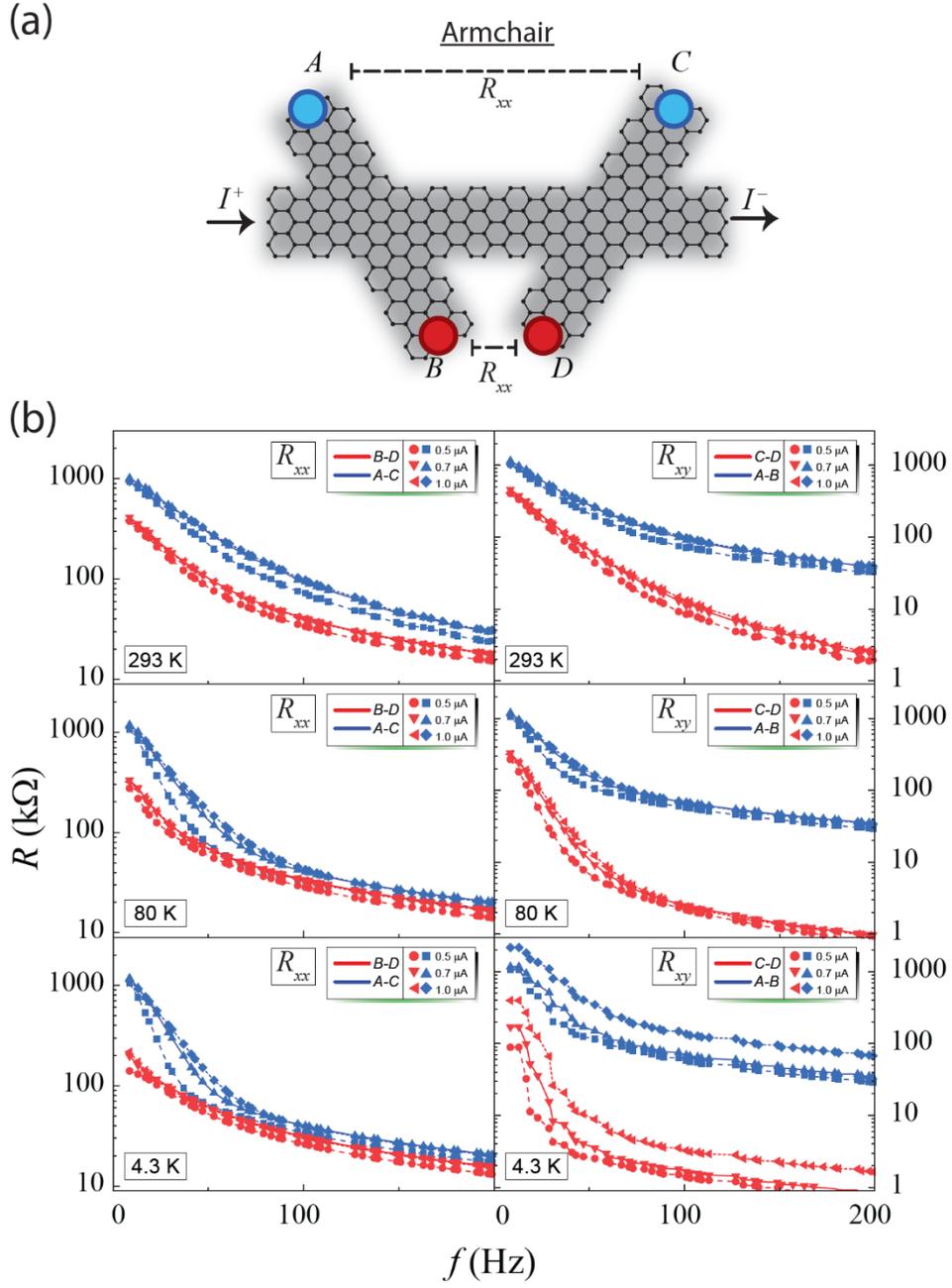

FIG. 3. (a) An illustration shows the general appearance of the AC GNR device, along with markers used in various electrical measurements. (b) A panel array is shown containing semi-logarithmic plots of the resistances as a function of frequency, with each panel dedicated to a particular temperature (from among 293 K, 80 K, and 4.3 K) and measurement configuration ($R_{xx}$ or $R_{xy}$). Within each plot, six curves, three per side of device, show this frequency dependent behavior. Electrical currents of 0.5 μA, 0.7 μA, and 1.0 μA are used for all data.

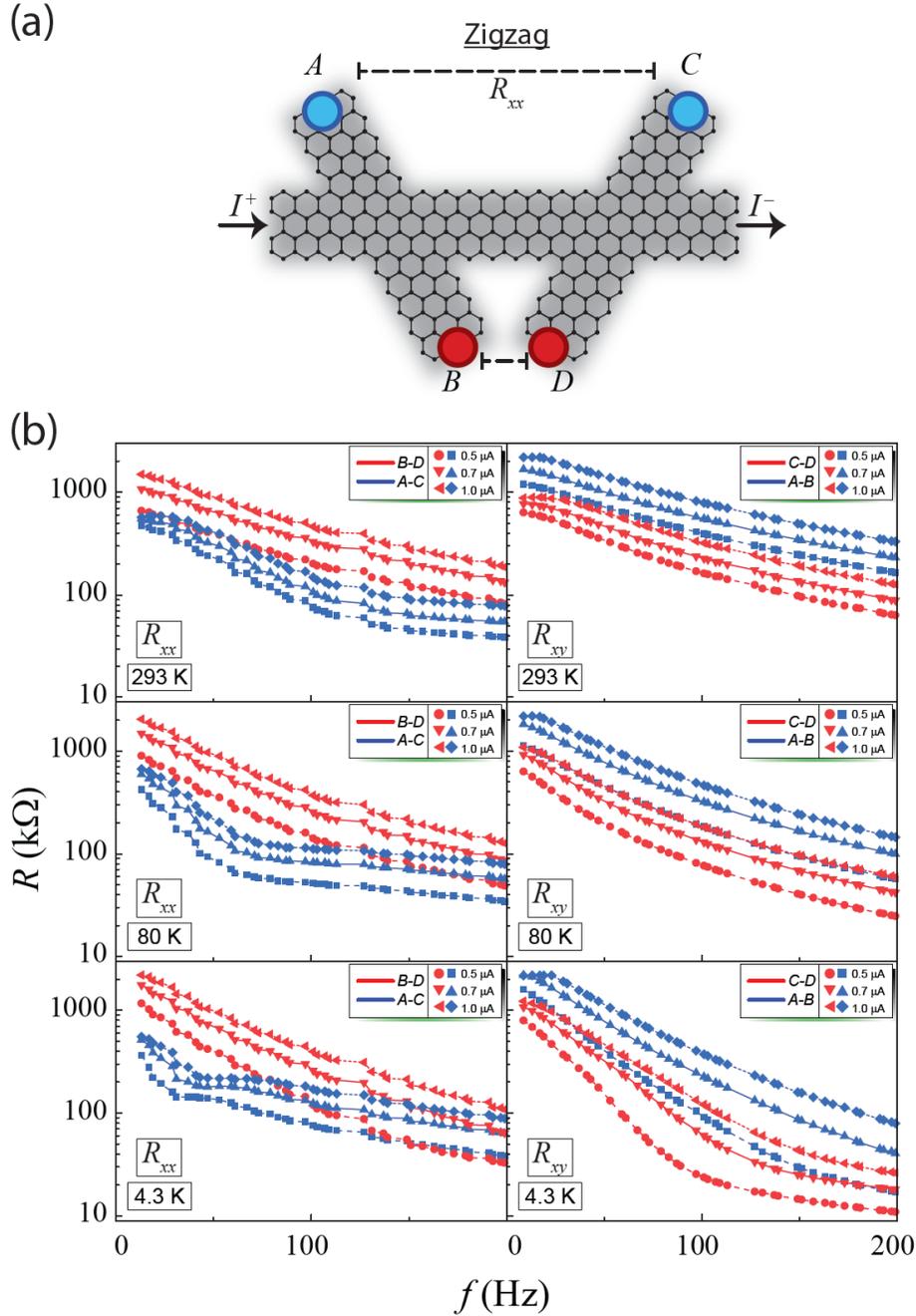

FIG. 4. (a) An illustration shows the general appearance of the ZZ GNR device, along with markers used in various electrical measurements. (b) A panel array is shown containing semi-logarithmic plots of the resistances as a function of frequency, with each panel dedicated to a particular temperature (from among 293 K, 80 K, and 4.3 K) and measurement configuration ($R_{xx}$ or $R_{xy}$). Within each plot, six curves, three per side of device, show this frequency-dependent behavior. Electrical currents of 0.5 μA, 0.7 μA, and 1.0 μA are used for all data.

Two interesting observations are made from these data. The first is a suggested distinction between the AC GNR and ZZ GNR devices' longitudinal resistance behaviors. The top and bottom trends are opposite in these two types, meaning that, for instance, the topside resistances are typically higher in AC GNR devices, whereas the opposite holds true in ZZ GNR devices. The second observation is that $R_{xx}$ appears to strongly exhibit the characteristic of being described by two exponential decay rates (two clear and conjoined linear behaviors in the logarithmic plots), whereas in the $R_{xy}$ case, this characteristic is less obvious, though still a possibility since no curve is obviously linear. These rates, which may be designated as resistance attenuation rates ($\Phi_1$ and $\Phi_2$, for low and high frequencies, respectively), can be extracted through a linear fit on the semi-logarithmic plot for lower frequencies (about 70 Hz and below) as well as higher frequencies (above 100 Hz). Ultimately, the equation used to describe the resistance attenuation behavior takes the form:

$$R(f) = R_0 + Ae^{-\alpha \frac{f}{\Phi_{1,2}}}$$

(1)

In Eq. 1, $\alpha$ is a simple factor based on the change of logarithmic base, and the constants $R_0$ and $A$ do not have a strong influence on the determined rates. Before continuing the analysis on these bases, it was crucial to justify their use. Two natural questions arise when looking to fit these data: (1) what form of model should be used? (2) how do we quantify the quality of that model since many models with different parameterizations may be called upon to describe the data?

To address the first question, one must understand the mathematical characterization of possible factors contributing to these observations of decreased resistance with frequency. Two main factors were considered: quantum capacitance and electron-phonon interactions. The quantum capacitance of a GNR has been described by the following [58]:

$$C_Q = \frac{e^2}{3ta\pi} \left( \frac{E - \frac{E_G}{2} + \frac{E_G}{2k_BT}}{\sqrt{E - \frac{E_G}{2} + \left(E - \frac{E_G}{2}\right)\frac{E_G}{k_BT}}} \right) \left(1 + e^{\left(\frac{E-E_F}{k_BT}\right)}\right)^{-1}$$

(2)

Equation 2 contains constants $e$ (elementary charge), $t$ (numerical value of hopping term in GNR [59]), $a$ (nearest neighbor distance), $E_G$ (band gap energy), $k_B$ (Boltzmann constant), $E_F$ (Fermi energy), and $T$ (temperature). In the limit of small-to-negligible band gap, this function behaves like an exponential decay as electron energy increases. At higher frequencies, therefore, the quantum capacitance could be expected to decrease, leading to a decrease in the effective resistance of the GNR devices. The second factor considered was the contribution from electron-phonon interactions. When an alternating current is applied to a GNR, the higher frequencies may provide additional energy to electrons such that they may overcome potential phonon-imposed energy barriers, thus resulting in a lower resistance [60]. Importantly, these electron-phonon interactions in GNRs have been describable in terms of Green's functions with terms that primarily depend on cotangent quantities, which themselves include exponential decays. These two major factors inspired the use of the exponential decay ansatz seen in Eq. 1.

To address the second question regarding model quality quantification, a Bayes factor analysis was employed to verify the appropriateness of the parameters used in the models based on Eq. 1. This analysis is provided in the Supplementary Material and can be summarized as showing that by calculating quantities like the marginal likelihood integral, the parameter covariance matrices, and the maximum likelihoods of these models, the ansatz of a two-term decay function is quantifiably more appropriate than a three-term (or more) decay function [61, 62].

An illustrated example of the two-term decay fitting procedure is shown in the inset of the top panel of Fig. 5. In both panels, every curve from Figures 3 and 4 was fit with a sloped line that, especially for the $R_{xx}$ data, visually characterized the rate of resistance decrease with frequency. The two resistance attenuation rates, $\Phi_1$ and $\Phi_2$, characterize lower frequencies (up to about 70 Hz) and higher frequencies (above approximately 100 Hz), respectively. The top panel of Fig. 5 shows the temperature dependence of these rates for both $R_{xx}$ and $R_{xy}$ in the AC GNR devices, whereas the bottom panel shows the same for the ZZ GNR devices. Overall, the rates do not appear to have a strong coupling to temperature, and in some cases, do not have an intuitive dependence at all.

What does become clear from these rate extractions is that there may be subtle differences suggested between the AC and ZZ GNR rates, and to extend on this knowledge, all devices were subjected to magnetic fields as high as 9 T. The orthogonal resistances were measured as a function of magnetic field and in both ramping directions (for $R_{xy}$ configurations AB and CD). Four different frequencies were used to allow one to compare $\Phi_1$ and $\Phi_2$ at 9 T and at 0 T. These sweeps are shown in Fig. 6 (a) and (c) for AC and ZZ GNR devices, respectively. To approximate $\Phi_1$ and $\Phi_2$ for the 9 T resistance data, a scaling ratio was calculated based on the change in resistance at a fixed magnetic field as a result of changing frequency. These estimations are shown in Fig. 6 (b) and (d), for AC and ZZ GNR devices, respectively. A comparison is graphically made between the two resistance attenuation rates that were extracted from the data in Fig. 6 (a). These comparisons highlight a very weak, and possibly no, dependence on magnetic field for both types of GNR, despite one type (AC) showing a less variable rate than the ZZ GNRs, as reflected by their error bars. The consistent behavior seen with the ZZ GNRs is a greater uncertainty in rate determinations, both at zero field and at 9 T. This manifests as an appearance of disagreement between the rates found for $R_{xx}$ and $R_{xy}$ measurements, a larger error bar from the spread of determined rates, and a generally higher $\Phi_1$ than the AC GNR case. Due to the high uncertainties, notably in the case of the ZZ GNRs, additional work is warranted to see if there exists any contribution from the inherent growth tendency of the ZZ GNR.

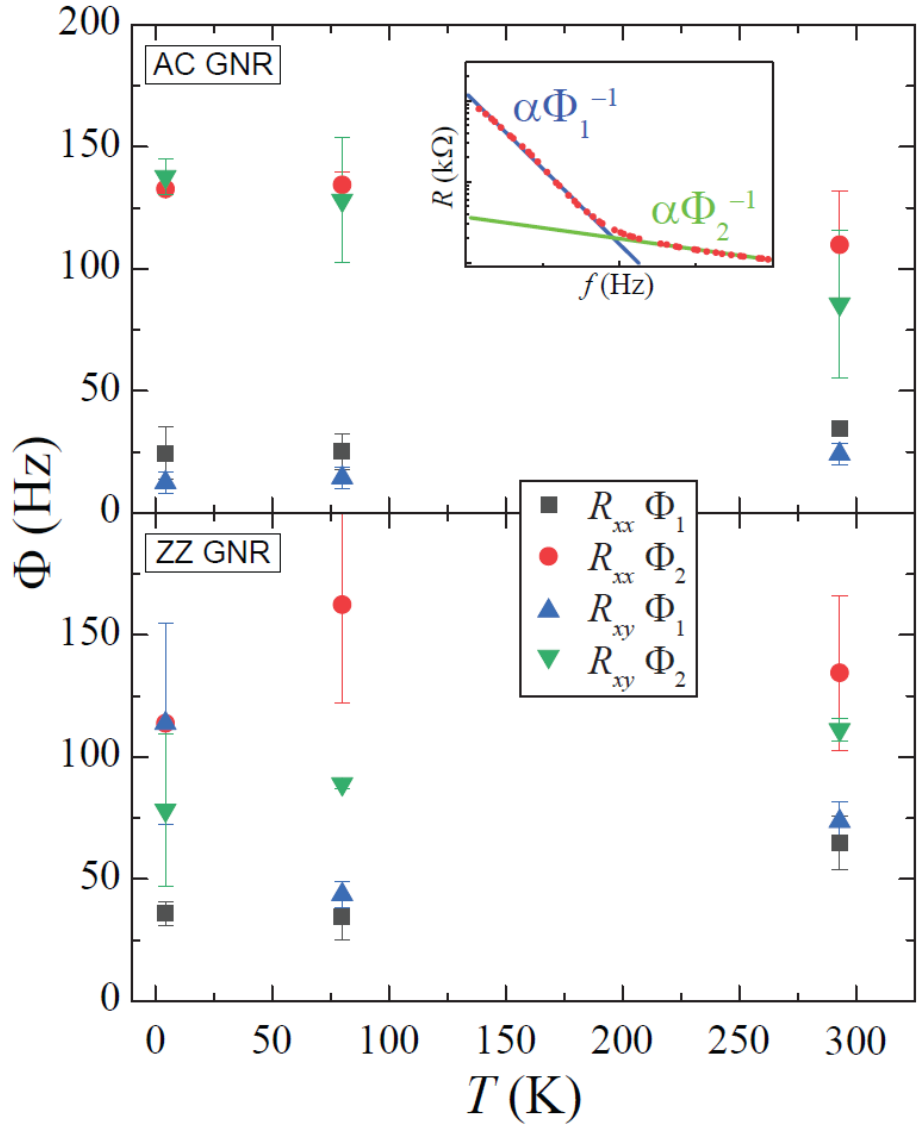

FIG. 5. Each curve from Figures 3 and 4 was analyzed with two linear fits that, especially for the $R_{xx}$ data, visually characterize the rate of resistance decrease with frequency. The two rates, $\Phi_1$ and $\Phi_2$, characterize lower frequencies (up to about 70 Hz) and higher frequencies (above approximately 100 Hz), respectively. The top panel shows the temperature dependence of these rates for both $R_{xx}$ and $R_{xy}$ in the AC GNR cases, along with an inset illustrating the general fitting procedure. The bottom panel shows the same data for the ZZ GNR cases. Error bars represent $1\sigma$ uncertainty of the average of all linear fits within each subset of data (labeled in the legend).

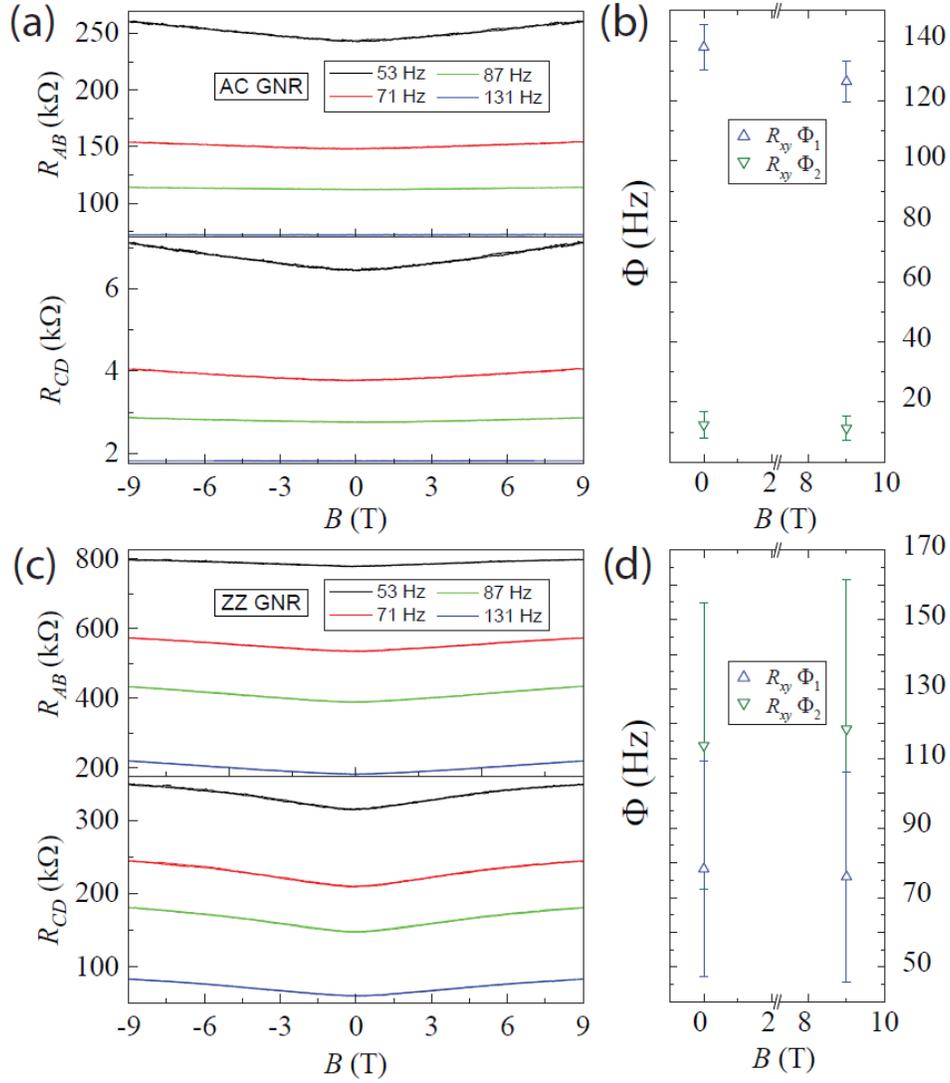

FIG. 6. (a) Magnetic field dependences were measured in both ramping directions for $R_{xy}$ configurations (AB and CD). Four different frequencies were used to later enable comparison of $\Phi_1$ and $\Phi_2$ with the zero-field counterparts. All data shown here are for AC GNR devices. (b) Magnetic field dependences are shown for the two resistance attenuation rates ($\Phi_1$ and $\Phi_2$) that are extracted from (a), highlighting a very weak, and possibly no, dependence on magnetic field. (c) Similar data are shown for the ZZ GNR devices as in (a), and $\Phi_1$ and $\Phi_2$ also show a lack of dependence on magnetic field in (d). Error bars represent 1σ uncertainty of the average of all linear fits within each subset of data (labeled in the legend).

In conclusion, growths of GNR devices about 50 nm wide on 4H-SiC have been demonstrated using PASG techniques. In this application, PASG promotes the deformation resistance of the slanted SiC sidewalls during the annealing process, allowing GNR growth to remain confined to the patterned SiC. The resistances of both AC and ZZ GNR devices were measured as a function of frequency for several temperatures, enabling one to extract rates of resistance attenuation that were observed in all data. These rates, along with the data from which they were extracted, demonstrate that alternating current and other techniques can be useful for distinguishing between AC and ZZ GNRs, especially when looking at frequencies under 70 Hz.

## DATA AVAILABILITY

The data that support the findings of this study are available from the corresponding author upon reasonable request.

## SUPPLEMENTARY MATERIAL

Supplementary Material at [URL] includes information for all fabrication processes and details regarding the Bayes factor analysis for modeling.

## ACKNOWLEDGMENTS

The work of C.-C. Y. and S. M. M at NIST was made possible by arrangement with C.-T. Liang of National Taiwan University. The authors would like to thank T. Mai, F. Fei, G. Fitzpatrick, and E. C. Benck for their assistance in the NIST internal review process. Work presented herein was performed, for a subset of the authors, as part of their official duties for the United States Government. Funding is hence appropriated by the United States Congress directly. The authors declare no competing interest or conflicts. The data that support the findings of this study are available from the corresponding author upon reasonable request.

Commercial equipment, instruments, and materials are identified in this paper in order to specify the experimental procedure adequately. Such identification is not intended to imply recommendation or endorsement by the National Institute of Standards and Technology or the United States Government, nor is it intended to imply that the materials or equipment identified are necessarily the best available for the purpose.

**Authors' Contribution statement: C.-C. Yeh**: Conceptualization, Formal Analysis, Investigation, Validation, Writing; **S. M. Mhatre**: Conceptualization, Formal Analysis, Investigation, Validation, Writing; **N. T. M. Tran**: Formal Analysis, Investigation, Validation, Writing; **H. M. Hill**: Formal Analysis, Investigation, Validation, Writing; **H. Jin**: Conceptualization, Investigation, Writing; **P.-C. Liao**: Formal Analysis, Investigation, Writing; **D. K. Patel**: Conceptualization, Formal Analysis, Writing; **R. E. Elmquist**: Conceptualization, Writing; **C.-T. Liang**: Conceptualization, Writing; and **A. F. Rigosi**: Conceptualization, Data Curation, Formal Analysis, Investigation, Methodology, Project Administration, Resources, Supervision, Validation, Writing

Measuring Epitaxial Graphene Nanoribbons with Low-Frequency Alternating Current

Ching-Chen Yeh[1,2,†], Swapnil M. Mhatre[1,3,†], Ngoc Thanh Mai Tran[1,4], Heather M. Hill[1], Hanbyul Jin[1,4], Pin-Chi Liao[3], Dinesh K. Patel[2,5], Randolph E. Elmquist[1], Chi-Te Liang[2], and Albert F. Rigosi[1]

[1]*Physical Measurement Laboratory, National Institute of Standards and Technology (NIST), Gaithersburg, MD 20899, United States*

[2]*Department of Physics, National Taiwan University, Taipei 10617, Taiwan*

[3]*Graduate Institute of Applied Physics, National Taiwan University, Taipei 10617, Taiwan*

[4]*Joint Quantum Institute, University of Maryland, College Park, Maryland 20742, United States*

[5]*Electricity Division, Physikalisch-Technische Bundesanstalt, 38116 Brunswick, Germany*

**GNR growth and fabrication procedures**

The details of the SiC substrate preparation, GNR growth, and additional fabrication steps are provided here.

The steps below are for SiC etching to create sidewalls for eventual GNR growth:

1. **Spin coating on bare SiC**

    a. LOR3A photoresist spun at 418.88 rad/s for 45 s

    b. Substrate baked at 180 °C for 5 min

    c. S1813 photoresist spun at 471.23 rad/s for 45 s

---

[†]These authors contributed equally to this work.

d. Substrate baked at 115 °C for 1 min

2. **UV Exposure**

    a. Use photolithography tool MLA150 at 375 nm with 90 mJ/cm$^2$

3. **Developing**

    a. CD 26A for 50 sec, followed by rinse with deionized water

4. **Metal deposition with e-beam deposition**

    a. Pd and Au (total respective thicknesses 10 nm and 150 nm) (respective deposition rates of 0.05 nm/s and 0.15 nm/s)

    b. Lift-off with PG remover at 80 °C for 1 hr

    c. Repeat with fresh PG remover for 5 min

5. **Dry etching of SiC**

    a. Reactive ion etching tool used with SF$_6$:O$_2$ (both gases have flow rate of 20 cm$^3$/min at standard temperature and pressure), 13.33 Pa, and 100 W for 90 s

    b. Note: this etches approximately 20 nm of SiC

6. **Removing etching mask**

    a. Place small patterns using S1813 to establish alignment marks for later steps

    b. Etch Au with a diluted solution of aqua regia and deionized water (1:1) for 2 min

    c. Remove photoresist S1813 with PG remover (80 °C for 10 min)

    d. Rinse substrate with isopropyl alcohol

The PASG technique and preparation for growth is summarized here:

1. **Sample cleaning**

a. Clean substrate with acetone, then isopropyl alcohol – each step is 10 min and includes sonication

   b. Place substrate in piranha bath at 120 °C for 10 min

   c. Hydrofluoric (HF) acid bath (49 % concentration) for 2 min (Step 2b immediately after)

2. **Spin-coating of polymer for growth assistance**

   a. Solution: 150 mL of isopropyl alcohol and 30 droplets (approximately 200 μL) AZ5214E

   b. Spin solution at 628.32 rad/s (with initial acceleration of 418.88 rad/s$^2$) for 30 s total process immediately after HF cleaning

3. **GNR growth (facing polished graphite disk)**

   a. 1200 °C for 30 min to induce sidewall sloping

   b. 1400 °C for 30 min to induce growth of GNRs

To <u>remove excess GNRs</u> outside the desired Hall bar region:

1. **Metal deposition for protective layer**

   a. Pd and Au (total respective thicknesses 10 nm and 30 nm) (respective deposition rates of 0.05 nm/s and 0.1 nm/s)

2. **Spin coating on bare SiC**

   a. LOR3A photoresist spun at 418.88 rad/s for 45 s

   b. Substrate baked at 180 °C for 5 min

   c. S1813 photoresist spun at 471.23 rad/s for 45 s

   d. Substrate baked at 115 °C for 1 min

3. **UV Exposure**

   a. Use photolithography tool MLA150 at 375 nm with 90 mJ/cm$^2$

4. **Developing**

    a. CD 26A for 50 sec, followed by rinse with deionized water

5. **Metal deposition with e-beam deposition**

    a. Cr (70 nm total thickness) (0.05 nm/s and 0.03 nm/s)

    b. Lift-off with PG remover at 80 °C for 1 hr

    c. Repeat with fresh PG remover for 5 min

6. **Dry etching of remaining Pd/Au and excess GNRs**

    a. Reactive ion etching tool used with Ar (40 cm³/min at standard temperature and pressure), 4 Pa, and 150 W for 10 min, then repeat for 3 min

7. **Removing etching mask**

    a. Use Cr etchant 1030 for 3 min

The next step places <u>electrical contact pads</u> on the GNR devices:

1. **Spin coating on bare SiC**

    a. LOR3A photoresist spun at 418.88 rad/s for 45 s

    b. Substrate baked at 180 °C for 5 min

    c. S1813 photoresist spun at 471.23 rad/s for 45 s

    d. Substrate baked at 115 °C for 1 min

2. **UV Exposure**

    a. Use photolithography tool MLA150 at 375 nm with 90 mJ/cm$^2$

3. **Developing**

    a. CD 26A for 50 sec, followed by rinse with deionized water

4. **Metal deposition with e-beam deposition**

a. Pd and Au (total respective thicknesses 10 nm and 200 nm) (respective deposition rates of 0.05 nm/s and 0.15 nm/s)

   b. Lift-off with PG remover at 80 °C for 1 hr

   c. Repeat with fresh PG remover for 5 min

   d. Note: this leaves part of the electrical contacts formed. The contact pads for wire bonding are listed next.

5. **Spin coating on bare SiC**

   a. LOR3A photoresist spun at 418.88 rad/s for 45 s

   b. Substrate baked at 180 °C for 5 min

   c. S1813 photoresist spun at 471.23 rad/s for 45 s

   d. Substrate baked at 115 °C for 1 min

6. **UV Exposure**

   a. Use photolithography tool MLA150 at 375 nm with 90 mJ/cm$^2$

7. **Developing**

   a. CD 26A for 50 sec, followed by rinse with deionized water

8. **Wire bonding metal deposition**

   a. Ti and Pt (total respective thicknesses 10 nm and 250 nm) (respective deposition rates of 0.05 nm/s and 0.1 nm/s)

   b. Lift-off with PG remover at 80 °C for 1 hr

Repeat with fresh PG remover for 5 min.

**Bayes factor analysis to validate models used**

It is through the calculation of the marginal likelihood integral (MLI) for each model that the Bayes factor, or ratio of MLIs [1], will better quantify the appropriateness of each of the fits.

One must first define the MLI, where *n* is the number of model parameters, *L<sub>max</sub>* is the maximum likelihood [2], **Cov<sub>p</sub>** is the parameter covariance matrix, and Δ*p* is the parameter value range [1]:

$$MLI = (2\pi)^{n/2} L_{max} \frac{\sqrt{det\ \mathbf{Cov_p}}}{\prod_{i=1}^{n} \Delta p_i}$$

(S1)

As mentioned earlier, the MLI characterizes the appropriateness of a fit to its corresponding dataset, and the Bayes factor emerges when the ratio between the MLIs for two different models is taken [1]. Typically, when the Bayes factor is calculated, the quality of the model represented in the numerator of the ratio is being compared to the model in the denominator of the ratio. Therefore, the goal is to assess the two-term decay model with respect to the three-term decay model, bearing in mind that a Bayes factor over 100 (or, alternatively for large datasets, a logarithm of the Bayes factor greater than 5) indicates that the model of interest (two-term decay model) is quantifiably more appropriate than the model to which it is compared (three-term decay model). A Bayes factor closer to 1 or lower suggests that the model of interest is not very appropriate to use in lieu of its counterpart.

When using several sets of data from the main text (Figure 6), this calculation yields a strong support for justifying the use of a two-term decay model as opposed to a three-term decay model (that is, the logarithm of the Bayes factor was much greater than 5).

REFERENCES

[1] D. J. Dunstan, J. Crowne, A. J. Drew, Sci. Rep. **12**, 993 (2022).

[2] D. J. C. MacKay, Neural Comput. **4**, 448 (1992).